\documentclass[preprint,numbered]{revtex4-1}

\usepackage{amssymb,amsmath}
\usepackage{hyperref}
\usepackage{bm}
\usepackage{bm}
\usepackage{graphicx}
\usepackage[title]{appendix}
\usepackage{lineno}
\usepackage{listings}

\usepackage{subfigure}
\usepackage{caption}

\usepackage{diagbox}

\newcommand{\matlab}{Matlab\textsuperscript{\textregistered}}

\begin{document}

\title{Meshfree and efficient modelling of swimming cells}
\author{Meurig T. Gallagher}
\email{m.t.gallagher@bham.ac.uk}
\author{David J. Smith}
\email{d.j.smith@bham.ac.uk}
\affiliation{School of Mathematics, University of Birmingham, Edgbaston, Birmingham, B15 2TT, UK}

\begin{abstract}
Locomotion in Stokes flow is an intensively-studied problem because it describes important biological phenomena such as the motility of many species' sperm, bacteria, algae and protozoa. Numerical computations can be challenging, particularly in three dimensions, due to the presence of moving boundaries and complex geometries; methods which combine ease-of-implementation and computational efficiency are therefore needed. A recently-proposed method to discretise the regularised Stokeslet boundary integral equation without the need for a connected `mesh' is applied to the inertialess locomotion problem in Stokes flow. The mathematical formulation and key aspects of the computational implementation in \matlab/GNU Octave are described, followed by numerical experiments with biflagellate algae and multiple uniflagellate sperm swimming between no-slip surfaces, for which both swimming trajectories and flow fields are calculated. These computational experiments required minutes of time on modest hardware; an extensible implementation is provided in a github repository. The nearest neighbour discretisation dramatically improves convergence and robustness, a key challenge in extending the regularised Stokeslet method to complicated, three dimensional, biological fluid problems.
\end{abstract}

\maketitle

\section{Introduction}
Inertialess locomotion in Stokes flow describes the motility of many types of sperm, bacteria, algae and protozoa. This topic has received extensive attention from mathematical modellers, starting with the classic work of Taylor \cite{taylor1951} and continuing to the present day \cite{keaveny2013,simons2015,ishimoto2017}. From the early work into the swimming of sea-urchin spermatozoa \cite{gray1955}, to investigations into the orientation of biflagellates in shear flows \cite{omalley2012}, there has been a lot of interest into modelling biological swimmers. This interest has been extended recently towards understanding and developing novel microswimmers. Topical examples of these involve studies into the microscale flow dynamics of ribbons and sheets \cite{montenegro201}, and the modelling of self-propelling toroidal swimmers based on the hypotheses of Taylor and Purcell \cite{huang2017}, as well as the study of phoretic toroidal swimmers \cite{schmieding2017}. Such works have the potential to enable the use of targeted drug delivery, amongst other things, through being able to guide microswimmers through complex biological environments \cite{montenegro2018}, and improve diagnostics and management of male infertility by analysis of imaging data.

Of particular recent interest is the collective behaviours of microswimmers. The differences in these behaviours appear to have significant biological implications, an example of which is the collective swimming of bovine sperm in the presence of viscoelasticity, behaviour which is not apparent in a purely viscous fluid \cite{tung2017}. Other species of sperm exhibit collective behaviours which impact both swimming and the ability to effectively fertilise the egg, some species of opossum sperm are often seen swimming as a cooperative pair \cite{cripe2016}. In addition to collective behaviours, the effects of interactions with other particles and/or boundaries have been recently shown to create interesting dynamics \cite{simons2014,lushi2017,shum2017,zottl2017}.

While each of the models presented above are in some sense idealised, the ability to further reduce detailed swimmer models to simplified representations provides the opportunity for extracting significant scientific information which may not be accessible otherwise. Such models allow for creation of a coarse-grained representation of a swimmer \cite{ishimoto2017} reducing complex behaviour into a set of swimming `modes' and their associated limit cycles. Detailed fluid dynamic modelling can also allow for calculation of parameters for continuum models \cite{pedley1990} and give understanding of hidden aspects of swimmers' characteristics such as energy transport along a flagellum \cite{gaffney2011} or internal moment generation \cite{brokaw1971}, as well as providing insight into the exact mechanisms for the collective swimming behaviours mentioned above.

Numerical methods are generally required to model finite amplitude motions, wall effects and swimmer-swimmer interactions. A range of numerical approaches exist, with perhaps the most extensively-studied being those based on singular, or regularized singular solutions of the Stokes flow equations, specifically resistive force theory \cite{gray1955}, slender body theory \cite{higdon1979}, boundary integral methods \cite{phan1987}, and regularized Stokeslet methods \cite{cortez2005,gillies2009,shankar2015,rostami2016}. These techniques remove the need to mesh the volume of fluid, requiring only the solution of integral equations formulated on the surface of the swimming body/bodies and lines such as cilia and flagella, reducing both the cost of meshing/remeshing a continually moving domain, and the number of degrees of freedom of the resulting linear system. Other techniques can be used to perform computational analysis of swimmers, such as the use of the force coupling method to investigate the dynamics of suspensions of up to $1000$ swimmers \cite{schoeller2018}, and the immersed boundary method \cite{peskin2002} for understanding the role of fluid elastic stress on flagellar swimming \cite{li2017}.

As reviewed recently \cite{smith2018}, regularized Stokeslet methods have the further major advantage of removing the need to evaluate weakly-singular surface integrals, and enabling slender bodies such as cilia and flage

To improve on the computational efficiency of the regularized Stokeslet method while retaining most of its simplicity of implementation, a method was proposed by Smith \cite{smith2018}, involving taking a coarser discretisation for the unknown traction than that used for numerical quadrature of the kernel, enabled by the use of nearest-neighbour discretisation. The method proved significantly more accurate for significantly lower computational cost, potentially enabling more complex and realistic problems to be investigated with given computational resources.

In this article we generalise the nearest-neighbour discretisation three dimensional regularized Stokeslet method to inertialess locomotion, in particular focusing on uniflagellate pushers modelling human sperm and a model biflagellate. In section~\ref{sec:fsp} we will briefly review the mathematical definition of the inertialess free-swimming problem in the boundary integral formulation. In section~\ref{sec:nnd} we implement the nearest-neighbour discretisation of the free-swimming problem with a single swimmer in an unbounded fluid. We then formulate the task of tracking the trajectory of the cell as an initial-value problem. The discretisation is then generalised in section~\ref{sec:generalisation} to incorporate rigid boundaries and multiple swimming cells. Finally, in section~\ref{sec:results} we present the results of numerical experiments with uniflagellate and biflagellate swimmers, and in section~\ref{sec:discussion} we discuss the method and further practical applications. Key aspects of the implementation in \matlab/GNU Octave are given, and a \verb@github@ repository is provided with the full code necessary to generate the results in the report, as well as templates for applying the method to novel problems in very low Reynolds number locomotion.

\section{The free-swimming problem}\label{sec:fsp}
The dynamics of a Newtonian fluid at very low Reynolds numbers, associated with locomotion of cells, is described by the Stokes flow equations. The dimensionless form of the equations is,
\begin{equation}
-\bm{\nabla}p+\nabla^2\bm{u}=0, \quad \nabla \cdot \bm{u}=0,\label{eq:stokes}
\end{equation}
augmented with the no-slip, no-penetration boundary condition 
\(\bm{u}(\bm{X})=\dot{\bm{X}}\) for boundary points \(\bm{X}\), where overdot denotes time-derivative. We note here that, for the kinematic-driven problems in the present paper, the viscosity term has been non-dimensionalised out of the PDE; for a force-driven problem the viscosity term would appear in the dimensionless group (the sperm number). Initially we will consider a single swimmer in a three dimensional unbounded fluid which is stationary at infinity. Two classical problems in Stokes flow are the \emph{resistance problem} -- which involves calculating the force and moment on a rigid body made to translate and rotate in stationary fluid, and the \emph{mobility problem} -- which involves calculating the rigid body motion due to an imposed force and moment.

The free-swimming problem in Stokes flow is a variant of the mobility problem. Rather than -- or perhaps in addition to -- the body being driven by imposed forces, it  translates and rotates as a result of changing its shape. In this section we will briefly review this problem, which has been solved numerically in many previous studies, and introduce our notation.

As usual for the regularized Stokeslet method, the fluid velocity \(u_j\) at location \(\bm{x}\) (suppressing time-dependence) is approximated by a surface integral over the surface \(\partial D\) of the swimmer,
\begin{equation}
 u_i(\bm{x})  \approx- \frac{1}{8\pi}  \iint_{\partial D} S_{ij}^{\epsilon}(\bm{x},\bm{X})f_j(\bm{X})dS_{\bm{X}}. \label{eq:rsbi}
\end{equation}
The regularisation error associated with equation~\eqref{eq:rsbi} has been discussed previously \cite{cortez2005} and will not be reviewed here. In this paper we will treat the approximation as exact.
The surface of the body will undergo motions that may be described by a model formulated in a body frame -- for example a frame in which the head of the cell does not move. If the body frame coordinates are \(\bm{\xi}\), and the body frame is described by the matrix of basis vectors (equivalently a rotation matrix) \(\bm{B}=(\bm{b}^{(1)}|\bm{b}^{(2)}|\bm{b}^{(3)})\) and origin \(\bm{x}_0\) then the laboratory frame coordinates and velocities are,
\begin{align}
\bm{x}       & = \bm{x}_0+\bm{B}\cdot \bm{\xi}, \\
\dot{\bm{x}} & = \dot{\bm{x}}_0 + \dot{\bm{B}}\cdot \bm{\xi} + \bm{B}\cdot\dot{\bm{\xi}}.
\end{align}
Denoting the rigid body velocity and angular velocity of the frame by \(\bm{U}\) and \(\bm{\Omega}\) respective, we then have,
\begin{align}
\bm{x}       & = \bm{x}_0 + \bm{B}\cdot \bm{\xi}, \\
\dot{\bm{x}} & = \bm{U}   + \bm{\Omega}\times (\bm{x}-\bm{x}_0) + \bm{B}\cdot\dot{\bm{\xi}}.
\end{align}

Applying the condition \(\bm{u}(\bm{x})=\dot{\bm{x}}\) on \(\partial D\) in equation~\eqref{eq:rsbi} yields the regularized Stokeslet boundary integral equation,
\begin{equation}
- \frac{1}{8\pi} \iint_{\partial D} S_{ij}^{\epsilon}(\bm{x},\bm{X})f_j(\bm{X})dS_{\bm{X}} = \dot{x}_i, \quad \mbox{all} \quad \bm{x} \in \partial D, \label{eq:rsbie}
\end{equation}
where it is understood that repeated indices (such as \(j\) in the above) are summed over, and unrepeated indices (such as \(i\) in the above) range over \(\{1,2,3\}\).

If at time \(t\), the body frame origin \(\bm{x}_0\) and orientation \(\bm{B}\) are known, and a model is given for the swimmer shape \(\bm{\xi}\) and motion \(\dot{\bm{\xi}}\) in the body frame, then then unknowns of the problem are the surface traction \(\bm{f}(\bm{X})\) for \(\bm{X}\in \partial D\), the translational velocity \(\bm{U}\) and angular velocity \(\bm{\Omega}\). The problem is closed by augmenting equation~\eqref{eq:rsbie} with the force and moment balance equations; here we assume that the inertia and moment of inertia of the swimmer are negligible. The full problem is then given by,
\begin{equation}\label{eq:swimprob}
\begin{aligned}
- U_i - \epsilon_{ijk} \Omega_j (x_k-x_{0k}) - \frac{1}{8\pi} \iint_{\partial D} S_{ij}^{\epsilon}(\bm{x},\bm{X})f_j(\bm{X})dS_{\bm{X}},
                    & = B_{ij} \dot{\xi}_j  \quad \mbox{all} \quad \bm{x} \in \partial D, \\
\iint_{\partial D} f_i(\bm{X})dS_{\bm{X}}                       & = 0,  \\
\iint_{\partial D} \epsilon_{ikj}X_k f_j(\bm{X}) dS_{\bm{X}}    & = 0,  
\end{aligned}
\end{equation}
where \(\epsilon_{ijk}\) is the Levi-Civita symbol.

Numerical discretisation of the problem~\eqref{eq:swimprob} will in general involve \(N\) vector degrees of freedom for  the traction \(\bm{f}\), three unknowns for the components of the translational velocity \(\bm{U}\) and three unknowns for the components of the angular velocity \(\bm{\Omega}\), totalling \(3N+6\) scalar unknowns in total. Through numerical collocation, problem~\eqref{eq:swimprob} can be formulated as \(3N+6\) linear equations. In the next section we will describe a nearest-neighbour regularized Stokeslet discretisation of this problem.

\section{Nearest-neighbour discretisation}\label{sec:nnd}
\subsection{A single swimmer in an unbounded fluid}\label{sec:singleUnbounded}
The discretisation of the regularized Stokeslet method is discussed in detail in \cite{smith2018}; in brief we suggest that a good balance of ease-of-implementation and numerical efficiency can be achieved by discretising the integrals via a quadrature rule, with the key modification of using a finer discretisation for the rapidly-varying regularised Stokeslet and a coarser discretisation for the more slowly-varying traction. A simple way to achieve this is through nearest-neighbour interpolation of the traction. The resulting method contains the original and extensively-used method of Cortez and colleagues \cite{cortez2005} as the limiting case in which the discretisations are equal.

Replacing the integrals in problem~\eqref{eq:swimprob} with numerical quadrature yields the discrete problem,
\begin{equation}\label{eq:swimprobnystrom}
\begin{aligned}
 - U_i - \epsilon_{ij\ell} \Omega_j (x_k[m]-x_{0k})
 - \frac{1}{8\pi}     \sum_{q=1}^Q S_{ij}^\epsilon(\bm{x}[m],\bm{X}[q])  f_j(\bm{X}[q]) dS(\bm{X}[q])
 & = B_{ij} \dot{\xi}_j[m] , \\
 & \mbox{for} \quad m=1,\ldots, N, \\
 \sum_{q=1}^Q f_i(\bm{X}[q]) dS(\bm{X}[q])     & = 0, \\
 \sum_{q=1}^Q \epsilon_{ikj}X_k[q] f_j(\bm{X}[q]) dS(\bm{X}[q])
                                               & = 0,
\end{aligned}
\end{equation}
where \(dS(\bm{X}[q])\) denotes the quadrature weight associated with the local surface metric. 

The coarse traction discretisation will be denoted as \(\{\bm{x}[1],\ldots,\bm{x}[N]\}\) and the finer quadrature discretisation as \(\{\bm{X}[1],\ldots,\bm{X}[Q]\}\); the \(Q\times N\) nearest-neighbour matrix is then
\begin{equation}
\mathsf{\nu}[q,n]=\begin{cases} 1 \quad \mbox{if} \quad n = \underset{\hat{n}=1,\ldots,N}{\mbox{argmin}} \, |\bm{x}[\hat{n}]-\bm{X}[q]|, \\ 0 \quad \mbox{otherwise}. \end{cases}
\end{equation}
A subtlety here concerns the calculation of the nearest-neighbour matrix when dealing with time-evolving geometries, and in particular the case when different bodies approach closely. As an example consider the case of a biflagellate swimmer: as the flagellum gets close to the body there is the potential for a quadrature point on the body to have a nearest force point on the flagellum (or \textit{vice versa}) leading to incorrect calculation of the traction at these points. For rigid bodies this is easily solved by calculating the nearest-neighbour matrix carefully at a single time point, before the bodies closely approach, and then treating $ \nu $ as constant in time. Alternatively one can calculate a time-evolving $ \nu $ on a body-by-body basis, considering separately (for example) the discretisations of a flagellum, cell body and any boundaries.

Defining \(g_i[n]:=-f_i(\bm{x}[n])\); the nearest-neighbour interpolation of the traction then corresponds to \(-f_i(\bm{X}[q]) dS(\bm{X}[q]) \approx\sum_{n=1}^N \mathsf{\nu}[q,n] g_i[n] dS(\bm{x}[n])\). Applying this interpolation to problem~\eqref{eq:swimprobnystrom} yields, 
\begin{equation}
  \begin{aligned}
   \frac{1}{8\pi} \sum_{n=1}^N g_j[n] dS(\bm{x}[n]) \sum_{q=1}^Q S_{ij}^\epsilon(\bm{x}[m],\bm{X}[q]) \mathsf{\nu}[q,n] 
 & - U_i - \epsilon_{ijk} \Omega_j (x_k[m]-x_{0k})  \\
                                            & = B_{ij} \dot{\xi}_j[m] ,  \quad \mbox{for} \quad m=1,\ldots, N, \\
   \sum_{n=1}^N g_j[n] dS(\bm{x}[n]) \sum_{q=1}^Q \delta_{ij} \nu[q,n] 
                                            & = 0, \\
   \sum_{n=1}^N g_j[n] dS(\bm{x}[n]) \sum_{q=1}^Q \epsilon_{ikj}X_k[q] \nu[q,n]
                                            & = 0. 
\end{aligned}\label{eq:swimprobnn}
\end{equation}
Computationally, problem~\eqref{eq:swimprobnn} corresponds to \(3N+3+3\) linear equations in \(3N+3+3\) scalar unknowns (\(F_j[n] := g_j[n] dS(\bm{x}[n])\) for \(n=1,\ldots,N\), followed by \(U_j\) and \(\Omega_j\)). These equations can be expressed in block form as,
\begin{equation}
  \begin{pmatrix}
    \,
    \\[0.5em]
    \, & A_{11}^S & \, & A_{12}^S & \, & A_{13}^S & \, & A_1^U  & A_1^\Omega   \\[3.9em]
    \, & A_{21}^S & \, & A_{22}^S & \, & A_{23}^S & \, & A_2^U  & A_2^\Omega   \\[3.9em]
    \, & A_{31}^S & \, & A_{32}^S & \, & A_{33}^S & \, & A_3^U  & A_3^\Omega   \\[2.0em]
    \, & A_1^F    & \, & A_2^F    & \, & A_3^F    & \, &        &              \\[0.5em]
    \, & A_1^M    & \, & A_2^M    & \, & A_3^M    & \, &        &              \\[0.2em]
  \end{pmatrix}
  \begin{pmatrix}
    F_1[1]  \\
    \vdots  \\[0.3em]
    F_1[N]  \\[0.7em]
    F_2[1]  \\
    \vdots  \\[0.3em]
    F_2[N]  \\[0.7em]
    F_3[1]  \\
    \vdots  \\[0.3em]
    F_3[N]  \\[0.5em]
    \bm{U}  \\[0.2em]
    \bm{\Omega}
  \end{pmatrix}
  =
  \begin{pmatrix}
    B_{1j}\dot{\xi}_j[1] \\
    \vdots               \\[0.2em]
    B_{1j}\dot{\xi}_j[N] \\[0.5em]
    B_{2j}\dot{\xi}_j[1] \\
    \vdots               \\[0.2em]
    B_{2j}\dot{\xi}_j[N] \\[0.5em]
    B_{3j}\dot{\xi}_j[1] \\
    \vdots               \\[0.2em]
    B_{3j}\dot{\xi}_j[N] \\[0.5em]
    \bm{0}               \\[0.2em]
    \bm{0}
  \end{pmatrix}
  ,
\end{equation}
where the blocks have entries given by,
\begin{equation}
  \begin{aligned}
    A_{ij}^S\{m,n\}   & = \frac{1}{8\pi}\sum_{q=1}^Q S_{ij}(\bm{x}[m],\bm{X}[q]) \nu[q,n] \quad & \mbox{for} \quad m,n=1,\ldots,N , \\
    A_i^U\{m,j\}      & = -\delta_{ij} \quad & \mbox{for} \quad m=1,\ldots, N, \\
    A_i^\Omega\{m,j\} & = -\epsilon_{ijk} (x_k[m]-x_{0k}) \quad & \mbox{for} \quad m=1,\ldots, N, \\
    A_j^F\{i,n\}      & = \delta_{ij} \sum_{q=1}^Q \nu[q,n] \quad & \mbox{for} \quad n=1,\ldots, N, \\
    A_j^M\{i,n\}      & = \epsilon_{ikj} X_k \sum_{q=1}^Q \nu[q,n]  \quad & \mbox{for} \quad n=1,\ldots, N,
  \end{aligned}
\end{equation}
and the velocity \(\bm{U}\) and angular velocity \(\bm{\Omega}\) are expressed as \(3\times 1\) column vectors. 

\subsection{Computing swimmer trajectories via an initial-value problem}
The position and orientation of a swimmer can be expressed as a position vector and a frame of basis vectors \(\bm{b}^{(j)}\). Given \(\bm{b}^{(1)},\bm{b}^{(2)}\), we then have \(\bm{b}^{(3)}=\bm{b}^{(1)}\times\bm{b}^{(2)}\) so it is sufficient to formulate the problem in terms of two basis vectors only, or six scalar degrees of freedom. Of course, this formulation still contains redundant information -- three Euler angles constrain precisely the body frame, however the basis vector approach is very straightforward to implement.

Noting that
\begin{equation}\label{eq:trajectoryprob}
\begin{aligned}
\dot{\bm{x}}_0     & = \bm{U}(\bm{x}_0,\bm{b}^{(1)},\bm{b}^{(2)},t), \\
\dot{\bm{b}}^{(j)} & = \bm{\Omega}(\bm{x}_0,\bm{b}^{(1)},\bm{b}^{(2)},t)\times {\bm{b}}^{(j)}, \quad j=1,2,
\end{aligned}
\end{equation} 
we may then formulate the calculation of trajectories as a system of \(9\) ordinary differential equations, where evaluation of the functions \(\bm{U}(\bm{x}_0,\bm{b}^{(1)},\bm{b}^{(2)},t)\) and \(\bm{\Omega}(\bm{x}_0,\bm{b}^{(1)},\bm{b}^{(2)},t)\) involves solving the swimming problem~\eqref{eq:swimprob}, for example via the discretisation~\eqref{eq:swimprobnn}.
The `outer' problem~\eqref{eq:trajectoryprob} can be solved using built-in functions such as \verb@ode45@ in \matlab or \verb@lsode@ in GNU Octave.

For practical purposes, when using a built-in initial value problem solver such as \verb@ode45@, the tractions \(f_i(\bm{X})\), required to compute the rate of energy dissipation and the flow field, may not be automatically available. To record this information, we may introduce the variable \(H_i(\bm{X},t)\), defined by,
\begin{equation}\label{eq:addivp}
  \begin{aligned}
    \dot{H}_i(\bm{x},t) & = f_i(\bm{x},t), \quad \bm{x}\in \partial D, \\
    H_i(\bm{x},0)       & = 0.
  \end{aligned}
\end{equation}
Augmenting the swimming problem~\eqref{eq:trajectoryprob} with equations~\eqref{eq:addivp} then yields an approximation to the force distribution available external to \verb@ode45@ by numerically differentiating \(H_i(\bm{x},t)\) with respect to time.

\section{Generalisation: boundaries and multiple swimmers}\label{sec:generalisation}
\subsection{Boundaries and fixed obstacles}
Mammalian sperm usually migrate and fertilise within a thin film of viscous fluid between opposed surfaces, and are typically imaged between a microscope slide and coverslip. Indeed, the major effect of boundaries on microswimmer flow fields has long been recognised \cite{liron1981}. Therefore it is important to take boundary effects into account in fluid dynamic simulations. The `Blakelet' and its regularized counterpart found by Ainley and colleagues \cite{ainley2008} (see also recent work by Cortez) is an elegant and efficient way to model a single infinite plane boundary; certain other geometrically simple situations possess similar fundamental solutions. However, it is important for full generality to take into account more complex boundary, and perhaps also fixed obstacles, present in the flow.

Representing the boundary by \(B\), the swimming problem becomes,
\begin{equation}\label{eq:swimprobbdry}
\begin{aligned}
 - U_i - \epsilon_{ijk} \Omega_j (x_k-x_{0k}) - \frac{1}{8\pi} \iint_{\partial D \cup B} S_{ij}^{\epsilon}(\bm{x},\bm{X})f_j(\bm{X})dS_{\bm{X}}
                    & = B_{ij} \dot{\xi}_j , \quad \mbox{all} \quad \bm{x} \in \partial D, \\
 - \frac{1}{8\pi} \iint_{\partial D \cup B} S_{ij}^{\epsilon}(\bm{x},\bm{X})f_j(\bm{X})dS_{\bm{X}}
                    & = \dot{x}_i , \quad \mbox{all} \quad \bm{x} \in B      ,  \\
  \iint_{\partial D} f_i(\bm{X})dS_{\bm{X}}                     & = 0        ,  \\
  \iint_{\partial D} \epsilon_{ijk}X_j f_k(\bm{X}) dS_{\bm{X}}  & = 0 .
\end{aligned}
\end{equation}

Numerically, we may represent the swimmer by the force points \(\{\bm{x}[1],\ldots,\bm{x}[N_s]\}\) and quadrature points \(\{\bm{X}[1],\ldots, \bm{X}[Q_s]\}\); the boundary is then discretised by the force points \(\{\bm{x}[N_s+1],\ldots,\bm{x}[N_s+N_b]\}\) and quadrature points \(\{\bm{X}[Q_s+1],\ldots, \bm{X}[Q_s + Q_b]\}\).
Nearest neighbour discretisation then leads to a system of the form,
\begin{equation}
  \begin{pmatrix}
    \,
    \\[0.5em]
    \, & A_{11}^S & \, & A_{12}^S & \, & A_{13}^S & \, & A_1^U  & A_1^\Omega   \\[3.9em]
    \, & A_{21}^S & \, & A_{22}^S & \, & A_{23}^S & \, & A_2^U  & A_2^\Omega   \\[3.9em]
    \, & A_{31}^S & \, & A_{32}^S & \, & A_{33}^S & \, & A_3^U  & A_3^\Omega   \\[2.0em]
    \, & A_1^F    & \, & A_2^F    & \, & A_3^F    & \, &        &              \\[0.5em]
    \, & A_1^M    & \, & A_2^M    & \, & A_3^M    & \, &        &              \\[0.2em]
  \end{pmatrix}
  \begin{pmatrix}
    \,
    \\[0.5em]
    F_1     \\[3.9em]
    F_2     \\[3.9em]
    F_3     \\[2.0em]
    \bm{U}  \\[0.5em]
    \bm{\Omega} \\[0.2em]
  \end{pmatrix}
  =
  \begin{pmatrix}
    \,
    \\[0.5em]
    V_1    \\[3.9em]
    V_2    \\[3.9em]
    V_3    \\[2.0em]
    \bm{0} \\[0.5em]
    \bm{0} \\[0.2em]
  \end{pmatrix}
  . \label{eq:blockSystem}
\end{equation}
The blocks have entries given by,
\begin{equation}
    \begin{aligned}
    A_{ij}^S\{m,n\}   & = \frac{1}{8\pi}\sum_{q=1}^Q S_{ij}(\bm{x}[m],\bm{X}[q]) \nu[q,n] \quad \mbox{for} \quad m,n=1,\ldots,N , \\
    A_i^U\{m,j\}      & =
    \begin{cases}
      -\delta_{ij} \quad & \mbox{for} \quad m=1,\ldots, N_s, \\
      0                   & \mbox{for} \quad m=N_s+1,\ldots, N_s+N_b,
    \end{cases}
    \\
    A_i^\Omega\{m,j\} & =
  \begin{cases}
    -\epsilon_{ijk} (x_k[m]-x_{0k}) \quad & \mbox{for} \quad m=1,\ldots, N_s, \\
    0                               & \mbox{for} \quad m=N_s + 1,\ldots, N_s+N_b,
  \end{cases}
  \\
    A_j^F\{i,n\}      & = \begin{cases}
      \delta_{ij} \sum_{q=1}^Q \nu[q,n] \quad & \mbox{for} \quad n=1,\ldots, N_s, \\
      0            \quad & \mbox{for} \quad n=N_s+1,\ldots, N_s+N_b,
                          \end{cases} \\
    A_j^M\{i,n\}      & = \begin{cases}
      \epsilon_{ikj} X_k \sum_{q=1}^Q \nu[q,n]  \quad & \mbox{for} \quad n=1,\ldots, N_s, \\
      0            \quad & \mbox{for} \quad n=N_s+1,\ldots, N_s+N_b,
                          \end{cases}
  \end{aligned}
\end{equation}
where the total number of force unknowns is $N = N_s + N_b$, the symbols \(F_j\) denote \((N_s+N_b)\times 1\) vectors of scalar unknowns \(F_j[1],\ldots,F_j[N_s+N_b]\), and the right hand sides are given by,
\begin{equation}
  V_i[n] =
  \begin{cases}
      B_{ij}\dot{\xi}_j[n] , \quad & \mbox{for} \quad n=1,\ldots, N_s, \\
      0                    , \quad & \mbox{for} \quad n=N_s+1, \ldots, N_s+N_b.
    \end{cases}
\end{equation}

\subsection{Multiple swimmers}\label{sec:multiple}
The last situation we will consider is where there are multiple swimmers --- which are not necessarily discretised by equal size sets --- as well as a boundary. The numerical discretisation is somewhat more complicated, and so we modify our notation in an attempt to make the implementation more interpretable.
Suppose that we now have \(N_{sw}\) swimmers, described by collocation points with \(i\)th components \(x_i^{(1)}[\cdot],\ldots, x_i^{(N_{sw})}[\cdot]\), their translational and angular velocities being denoted \(U_i^{(1)},\ldots, U_i^{(N_{sw})}\) and \(\Omega_i^{(1)},\ldots, \Omega_i^{(N_{sw})}\); the boundary points will be denoted by the array \(x_i^{(b)}[\cdot]\).
The discretisation will follow the ordering convention,
  \begin{equation}
  \mathsf{x} = \begin{pmatrix} 
  		\mathsf{x}_1\{\cdot\},\quad \mathsf{x}_2\{\cdot\},\quad\mathsf{x}_3\{\cdot\} 
  		\end{pmatrix}^T, \quad\text{with}\quad \mathsf{x}_i\{\cdot\} = \begin{pmatrix} x_i^{(1)}[\cdot],\quad \hdots,\quad x_i^{(N_{sw})}[\cdot],\quad x_i^{(b)}[\cdot] 
  		\end{pmatrix}^T,
  \end{equation}
  which is inherited by the right hand side velocities and the force discretisation. If the number of force points associated with swimmer \(r\) is \(N_s(r)\), and the number of force points associated with the boundary is \(N_b\), then the number of vector force unknowns is \(N_f=\sum_{r=1}^{N_{sw}} N_s(r)+N_b\). The size of \(\mathsf{x}\) is then \(3N_f\) and the total number of scalar degrees of freedom in the system is \(3N_f + 6N_{sw}\). We will define the index \(\iota(r)\) to be the location of the \(r\)th swimmer in the \(\mathsf{x}_i\) vector, with \(\iota(1)=1\) and \(\iota(r)=\sum_{\gamma=1}^{r-1} N_{s}(\gamma)\) for \(1<r\leqslant N_{sw}\). 

  The quadrature points may be denoted \(\bm{X}[1],\ldots, \bm{X}[Q]\) as previously; the Stokeslet matrix is then constructed as,
  \begin{equation}
    A_{ij}^S\{\alpha,\beta\} = \frac{1}{8\pi} \sum_{q=1}^Q S_{ij} (\mathsf{x}_i[\alpha], \bm{X}[q] ) \nu[q,\beta],    \quad \mbox{for} \quad \alpha, \beta = 1,\ldots,N_f.
  \end{equation}
  To construct the remaining blocks, we introduce the notation \(\mathbf{1}^{(n)}\) to be the column vector of length \(n\) with every entry equal to \(1\) and \(\mathbf{0}^{(m\times n)}\) to be the \(m\times n\) matrix of zeros. We also define the \(N_f\times N_{sw}\) matrices,
  \begin{equation}
    \tilde{x}_i\{\cdot,\cdot\} = \begin{pmatrix} \, & & \\[0.5em]  x_i^{(1)}[\cdot]-x_{0i}^{(1)}   &    &    \\[2.0em]   &  \ddots  &    \\[2.0em]  &    &   x_i^{(N_{sw})}[\cdot]-x_{0i}^{(N_{sw})}  \\[3.0em]  & \, \mathbf{0}^{(N_b\times N_{sw})} &  \\[3.0em]  \end{pmatrix}. 
  \end{equation}
Then,
  \begin{equation}
    A^U = I_3 \otimes \begin{pmatrix} \, & & \\[0.5em]  -\mathbf{1}^{(N_s(1))}  &      &     \\[2.0em]    & \ddots &   \\[2.0em]    &    &  -\mathbf{1}^{(N_s(N_{sw}))}\\[3.0em]   & \, \mathbf{0}^{(N_b\times N_{sw})}  &  \\[3.0em]  \end{pmatrix},\quad\text{and}\quad
    A^\Omega = \begin{pmatrix}    & \, & \\[0.5em] &  - \tilde{x}_3\{\cdot,\cdot\}   & \tilde{x}_2\{\cdot,\cdot\}  \\[2.0em] \tilde{x}_3\{\cdot,\cdot\}  &   & -\tilde{x}_1\{\cdot,\cdot\}  \\[2.0em]  -\tilde{x}_2\{\cdot,\cdot\}  & \tilde{x}_1\{\cdot,\cdot\} &  \\[2.0em]  \end{pmatrix},
  \end{equation}
  with \(\otimes\) denoting the Kronecker product.

Recalling that \(\nu[\cdot,\cdot]\) denotes the nearest-neighbour matrix, we define the \(N_s(r) \times 1\) column vectors,
\begin{equation}
  \begin{aligned}
    \lambda^{(r)}[\cdot] & = \sum_{q=1}^Q \nu[q,\iota(r):\iota(r+1)-1], \\
     \chi_j^{(r)}[\cdot] & = \sum_{q=1}^Q X_j(q) \nu[q,\iota(r):\iota(r+1)-1],
  \end{aligned}  
\end{equation}
and the \(N_{sw}\times N_f\) matrices,
\begin{equation}
  \tilde{\chi}_j \{\cdot,\cdot\} = \begin{pmatrix} \quad \chi_j^{(1)T}[\cdot] \quad &  &  &  \\   & \ddots  &  &  \quad \mathbf{0}^{(N_{sw}\times N_{b})} \quad \\  &  &  \quad \chi_j^{(N_{sw})T}[\cdot] \quad &  \end{pmatrix}.
\end{equation}
Then the \(3N_{sw} \times 3N_f\) blocks \(A^F\) and \(A^M\) are,
\begin{equation}
  \begin{aligned}
    A^F & = I_3 \otimes \begin{pmatrix} \quad \lambda^{(1)T}[\cdot] \quad & & & \\ & \ddots & & \quad \mathbf{0}^{(N_{sw}\times N_{b})} \quad \\ & & \quad \lambda^{(N_{sw})T}[\cdot] \quad & \end{pmatrix},  \\
    A^M & = \begin{pmatrix} \, & & \\  &  \quad -\tilde{\chi}_3 \quad & \quad \tilde{\chi}_2 \quad \\[1.0em] \quad \tilde{\chi}_3 \quad &   & \quad -\tilde{\chi}_1 \quad \\[1.0em] \quad -\tilde{\chi}_2 \quad  & \quad \tilde{\chi}_1 \quad \\ & & \, \end{pmatrix}.
  \end{aligned}
\end{equation}

Finally, denoting the orientation matrix of the \(r\)th swimmer by \(B_{ij}^{(r)}\) and its body frame waveform as \(\xi_j^{(r)}\), the terms of the right hand side take the form,
\begin{equation}
V_i = \begin{pmatrix} V_i^{(1)}[\cdot],\quad \hdots,\quad V_i^{(N_{sw})}[\cdot],\quad\mathbf{0}^{(N_{sw}\times 1)}\end{pmatrix}^T,
\end{equation}
where
\begin{equation}
  V_i^{(r)}[n] = B_{ij}^{(r)} \dot{\xi}_j^{(r)}[n].
\end{equation}
Now that we have defined \(A_{ij}^S\), \(A^U\), \(A^{\Omega}\), \(A^F\), \(A^M\) and \(V_i\), the \(3(N_f+2)\times 3(N_f+2)\) linear system is of the form given by equation~\eqref{eq:blockSystem}.

\section{Results and analysis}\label{sec:results}

We now turn our attention to the application of this method to two model problems: (1) a single biflagellate swimming in an infinite fluid, and (2) multiple sperm cells swimming between two boundaries. The implementation for both these model problems is provided in the associated github repository. After presenting the results for these swimming problems we will discuss the convergence of the method for the two types of swimmer provided and compare with the results obtained through the classic Nystr\"om discretisation (when the force and quadrature discretisations are the same).

\subsection{Biflagellate in an infinite fluid}
\label{sec:biflagellate}

\begin{figure}
  \centering
  \includegraphics[width=\textwidth]{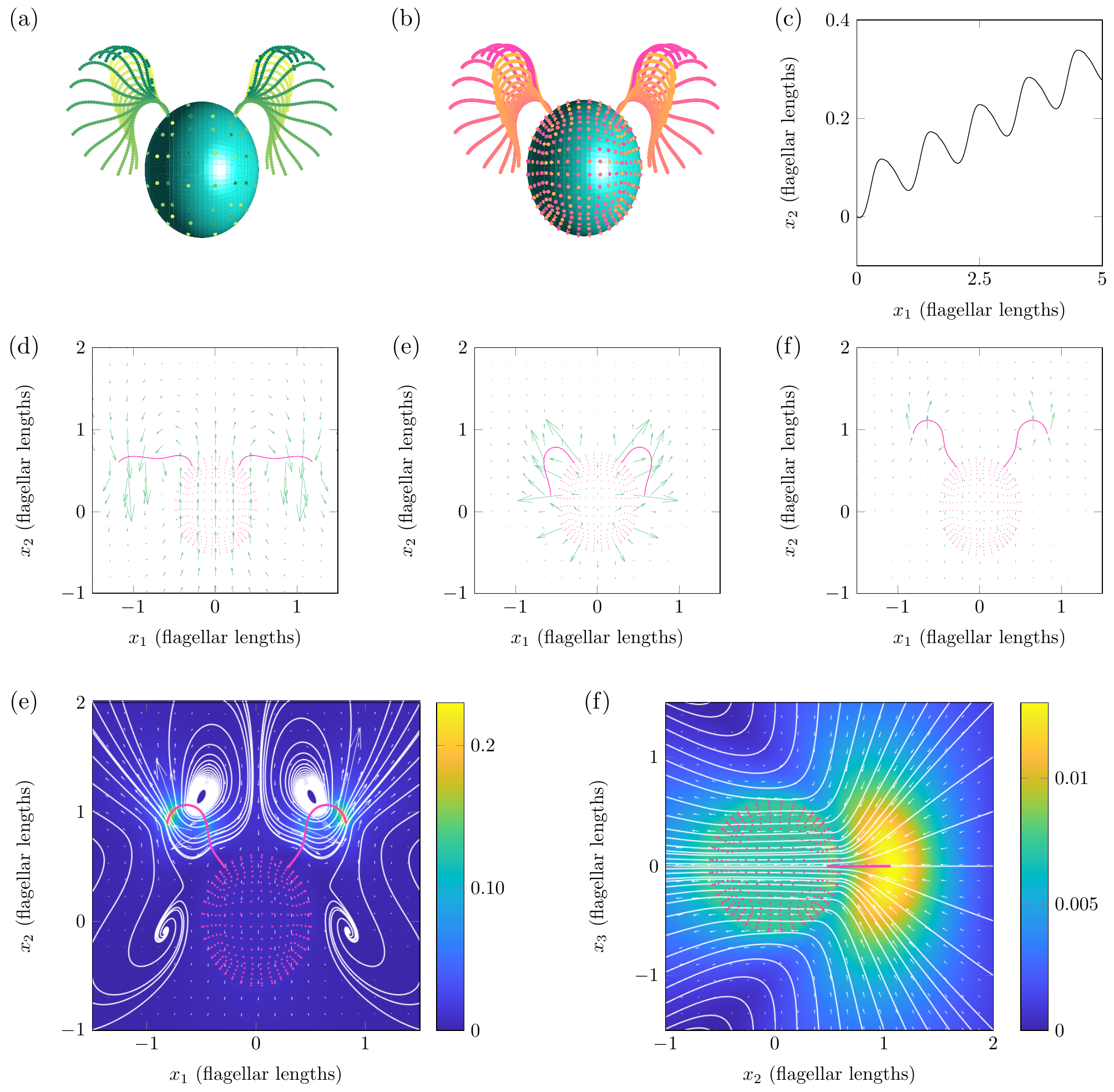}
  \caption{\footnotesize{Computational results for a free-swimming model biflagellate in an unbounded fluid, implemented with the script \texttt{GenerateSwimmingFigureChlamy.m} (a,b) Model biflagellate showing beat pattern, visualised via (a) force discretisation, (b) quadrature discretisation. (c) $x_2$ coordinate of free-swimming cell over five beat cycles, where positive $x_2$ is the overall swimming direction. (d,e,f) computed flow fields at (d) $t=2\pi/3$, (e) $t=4\pi/3$ and (f) $t=2\pi$ (three points during the beat cycle). (g,h) computed velocity profile with streamlines at $t = 0$, cross sections of the (g) $\left(x_1,x_2\right)$ and (h) $\left(x_2,x_3\right)$ planes respectively.}}
  \label{fig:chlamy}
\end{figure}

We will first apply the algorithm in section~\ref{sec:singleUnbounded} to model a biflagellate, superficially similar to various marine algae, swimming in an unbounded fluid. We model the beat pattern of the cell (figure~\ref{fig:chlamy}a) following Sartori et al \cite{sartori2016}, writing the flagellar tangent angle $ \psi $ in the form
\begin{equation}
	\psi\left(s,t\right) = \psi_0\left(s\right) - \psi_1\left(s\right)\cos{\left(t + \phi\left(s\right)\right)},
\end{equation}
where $ s $ and $ t $ are dimensionless arclength along the flagellum and time respectively. We find that choosing
\begin{equation}
	\psi_0\left(s\right) = -2.5 s,\quad \psi_1 = 0.7 + 0.15\sin\left(2\pi s\right),\quad \phi\left(s\right) = -2\pi s,\quad 0 \leq s \leq 1,
\end{equation}
provides a sufficiently representative test case for the computational algorithm. Of course a more realistic beat for a genuine biflagellate species such as \emph{Chlamydomonas reinhardtii} could be appended as required.

The two flagella are synchronised; for the force discretisation, 40 points are used to discretise each flagellum, and 96 points are used for the cell body, totalling \(176\) vector degrees of freedom (figure~\ref{fig:chlamy}). For the quadrature discretisation, \(400\) points are used for each flagellum, and \(600\) points for the cell body, giving a total of \(1400\) quadrature points (figure~\ref{fig:chlamy}b). The regularisation parameter is chosen as $ \epsilon = 0.25/20$ to represent the radius of the flagellum (scaled with flagellar length).

Results showing the displacement of the swimming cell are shown in figure~\ref{fig:chlamy}c, and the flow field at three points of the beat in figure~\ref{fig:chlamy}d, \ref{fig:chlamy}e and \ref{fig:chlamy}f. The latter calculation can be carried out in a `post-processing' step from the computed swimmer position, orientation and force distribution. To further visualise the flow we have included in figures \ref{fig:chlamy}g and \ref{fig:chlamy}h a selection of streamlines plotted over the fluid velocity. While the figures show a 2D projection, the computation is fully three-dimensional, and the instantaneous flow field on any (finite) subset of \(\mathbb{R}^3\) can be computed. The computation and creation of figure~\ref{fig:chlamy} required \(33.4\)~s on a desktop computer (2017 Lenovo Thinkstation P710; Intel(R) Xeon(TM) E5-2646 CPU @ 2.40GHz; 128GB 2400 MHz RDIMM RAM).

\subsection{Sperm between two opposed surfaces}
\label{sec:sperm}

\begin{figure}
  \centering
   \includegraphics[width=\textwidth]{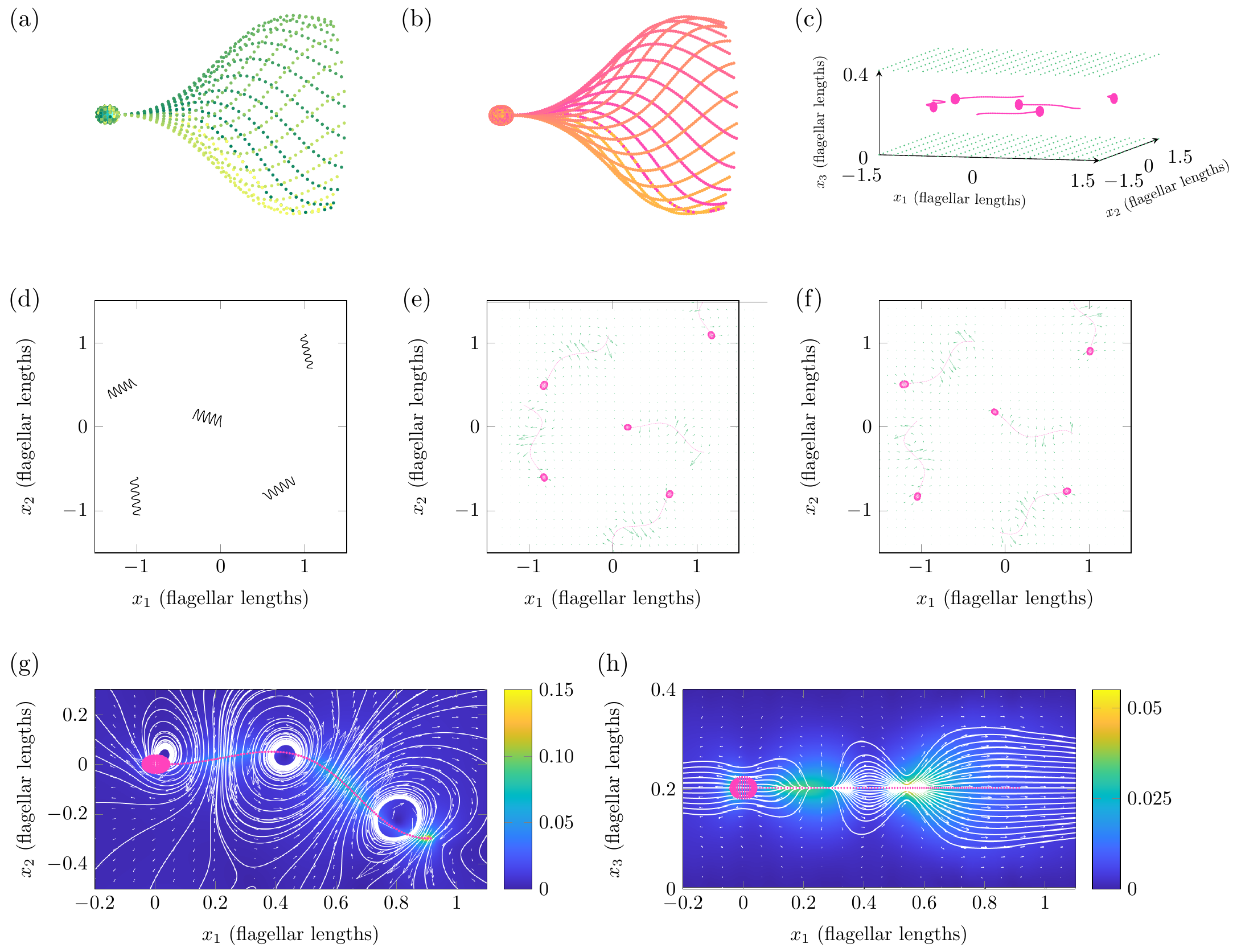}
   \caption{\footnotesize{Computational results for a group of free-swimming model sperm swimming midway between two opposed no-slip surfaces separated by \(0.4\) flagellar lengths, implemented with the script \texttt{GenerateSwimmingFigureSperm.m} (a,b) Model sperm showing beat pattern of Dresdner \& Katz \cite{dresdner1981}, visualised via (a) force discretisation, (b) quadrature discretisation. (c) Visualisation of sperm placed between the discretised boundaries (note that the sperm heads appear rounder than they actually are due to the aspect ratio chosen for plotting). (d) Trajectories of free-swimming cells over five beat cycles. (e,f) computed cell positions (with 5 randomly-perturbed beat cycles and head dimensions), and flow fields at (e) \(t=0\) and (f) \(t=2.5\)~cycles. (g,h) computed velocity profile with streamlines of a single sperm at \(t=0\), cross sections of the (g) $\left(x_1,x_2\right)$ and (h) $\left(x_1,x_3\right)$ planes respectively.}}\label{fig:sperm}
\end{figure}

We now turn our attention to the more general problem of section~\ref{sec:multiple} involving multiple swimming cells and boundaries. The computational domain contains two no-slip square surfaces with sides of length \(3L\), separated by a distance \(0.4L\), where \(L\) is the flagellar length (for human sperm typically \(L\approx 45\)~\(\mu\)m). The swimmer heads are ellipsoids with axes of length $0.044L$, $0.036L$ and $0.022L$. The flagellar movement is based on the classic planar `activated' beat of Dresdner \& Katz \cite{dresdner1981}; the sperm head (cell body) is a scalene ellipsoid. Figures~\ref{fig:sperm}a and \ref{fig:sperm}b show the beat pattern via the force and quadrature discretisations respectively. The force discretisation consists of \(136\) points per cell and \(480\) points for the boundary, totalling \(3480\) scalar degrees of freedom for a simulation with five cells. The quadrature discretisation consists of \(700\) points per swimmer and \(1920\) points for the boundary, totalling \(5420\) quadrature points. The regularisation parameter is chosen as $ \epsilon = 0.25/45$ to represent the radius of the flagellum (scaled with flagellar length).

The computation shown in figure~\ref{fig:sperm} involves tracking five cells each with slightly perturbed beat cycle and head morphology parameters, swimming mid-way between the no-slip boundaries described above (visualised in figure~\ref{fig:sperm}c), for five beat cycles. Figure~\ref{fig:sperm}d shows the cell trajectories, and figures~\ref{fig:sperm}d and \ref{fig:sperm}e show the cell positions, orientations and surrounding flow fields at two distinct time points. To further visualise the flow we have included in figures \ref{fig:sperm}g and \ref{fig:sperm}h a selection of streamlines plotted over the fluid velocity. The calculated dimensionfull swimmer velocity is $ \approx 43~\mu\mathrm{m s}^{-1} $, this is comparable to the results of Smith \textit{et al.} \cite{smith2009} who report a numerical calculation of the speed of a sperm with the same waveform, swimming at a distance $0.2$ flagellar lengths from a surface, as $\approx 42~\mu\mathrm{m s}^{-1}$. While the computation was more intensive than that described in the previous section, it was still easily within reach of the same computer, requiring $127~s$ of wall time.

\subsection{Convergence of the method with discretisation refinement}
\label{sec:convergence}

A practical refinement heuristic for assessing the convergence (with increased discretisation) of the nearest neighbour method is given by Smith \cite{smith2018}. For testing the convergence of the present swimming problems we denote the maximum discretisation spacings from \cite{smith2018} as
\begin{align}
	h_f = \max_{m=1,\hdots,N} \min_{\substack{n = 1,\hdots,N\\ n \neq m}} \left\lvert \bm{x}\left[m\right] - \bm{x}\left[n\right]\right\rvert,\quad h_q = \max_{p=1,\hdots,Q} \min_{\substack{q = 1,\hdots,Q\\ q \neq p}} \left\lvert \bm{X}\left[p\right] - \bm{X}\left[q\right]\right\rvert.
\end{align}

In the present work we note that we may have different discretisations for each swimmer, and indeed for each component of a single swimmer (the head and flagellum may be discretised differently for example). To this end we apply the existing convergence heuristic in stages as outlined in table \ref{tab:heuristic}. To measure the convergence we compare the straight line distance travelled over one full beat of the swimmer's flagellum. In contrast to the classical (Nystr\"{o}m) discretisation \cite{cortez2005}, there is no tight coupling between the regularisation parameter $\epsilon$ and the discretisation length scales \cite{smith2018}.  As a consequence of this we allow the choice of regularisation parameter $ \epsilon $ to be guided by the geometry of the swimmer (chosen here to be related to the dimensions of the flagellum).

\begin{table}[t]
	\caption{Heuristic for analysing the convergence of the results.}
	\label{tab:heuristic}
	\hrulefill
	\begin{enumerate}
		\item Generate an initial force and quadrature discretisation for the swimmer head $ h_f^H $ and $ h_q^H $.
		\item Assess convergence by the heuristic in \cite{smith2018} through varying the flagellar discretisations $ h_f^F $ and $ h_q^F $.
		\item Generate a more refined head quadrature discretisation by halving $ h_q^H $ and repeat step 2.
		\item Generate a more refined head force discretisation by halving $ h_f^H $ and repeat step 2.
		\item Repeat steps 3 and 4 until a suitable level of convergence is reached.
	\end{enumerate}
	\hrulefill
\end{table}

\begin{table}[!htp]
	\footnotesize
	\centering
	\caption{New nearest-neighbour convergence results: Straight line distance traveled by a single biflagellate swimmer, as described in \S\ref{sec:biflagellate}, after one complete flagellar beat cycle. Here, the discretisation for the cell head is fixed with $ N^H = 96 $ and $ Q^H = 600 $ force and quadrature points respectively. The number of points discretising the flagellum have been chosen following the convergence algorithm in table \ref{tab:heuristic}, with the regularisation parameter $ \epsilon = 0.25/20 $ being the ratio between flagellar radius and length. When $ N^F > Q^F $ a singular linear system is formed, this is denoted by the entry `NaN' (`not-a-number'). (a) Distance traveled by the swimmer, (b) and (c) show the percentage change in this distance when halving $ h_f^F $ and $h_q^F $ respectively.}
	\label{tab:chlamy1}
	\subtable[\ Distance traveled in multiples of $ \left(\text{flagellar length}\right)\cdot 10^{-2} $]{
		\begin{tabular}{ l l l l l| l l l l }
			         &        &          &       & \multicolumn{1}{r|}{$Q^H$}  & $600$     & $600$     & $600$     & $600$       \\		
			         &        &          &       & \multicolumn{1}{r|}{$h_q^H$}& $0.1137$  & $0.1137$  & $0.1137$  & $0.1137$   \\
			         &        &          &       &  \multicolumn{1}{r|}{$Q^F$} & $100$     & $200$     & $400$     & $800$      \\
			 DOF     & $N^H$  & $h_f^H$  & $N^F$ & \diagbox{$h_f^F$}{$ h_q^F$} & $0.09767$ & $0.04859$ & $0.02424$ & $0.01210$ \\
			\hline
			$ 528 $ & $ 96 $ & $0.2272$ & $ 40 $  & $ 0.2479 $                & $ 5.521 $ & $ 5.521 $ & $ 5.526 $ & $ 5.527 $ \\
			$ 768 $ & $ 96 $ & $0.2272$ & $ 80 $  & $ 0.1224 $                & $ 5.503 $ & $ 5.501 $ & $ 5.498 $ & $ 5.498 $ \\
			$ 1248 $ & $ 96 $ & $0.2272$ & $ 160 $ & $ 0.06082 $               & NaN       & $ 5.504 $ & $ 5.503 $ & $ 5.502 $ \\
			$ 2208 $ & $ 96 $ & $0.2272$ & $ 320 $ & $ 0.03032 $               & NaN       & NaN       & $ 5.487 $ & $ 5.432 $
		\end{tabular}
	}
	
	\subtable[\ Percentage change in distance traveled when halving $ h_f^F $]{
		\begin{tabular}{ l | l l l l l}
			\diagbox{$h_f^F$}{$ h_q^F$} & $0.09767$ & $0.04859$   & $0.02424$   & $0.01210$ \\
			\hline
			$ 0.1224 $                  & $ 0.32\%$ & $ 0.37\% $  & $ 0.49\% $  & $ 0.52\% $  \\
			$ 0.06082 $                 &           & $ 0.06\% $  & $ 0.08\% $  & $ 0.08\% $\\
			$ 0.03032 $                 &           &             & $ 0.29\% $  & $ 1.28\% $ 
		\end{tabular}
	}\qquad\qquad	
	\subtable[\ Percentage change in distance traveled when halving $ h_q^F $]{
		\begin{tabular}{ l | l l l l}
			\diagbox{$h_f^F$}{$ h_q^F$} & $0.04859$  & $0.02424$   & $0.01210$\\
			\hline
			$ 0.2479 $                  & $ 0.01\% $ & $ 0.08\% $  & $ 0.02\% $ \\
			$ 0.1224 $                  & $ 0.03\% $ & $ 0.05\% $  & $ 0.01\% $  \\
			$ 0.06082 $                 &            & $ 0.02\% $  & $ 0.01\% $  \\
			$ 0.03032 $                 &            &             & $ 1.00\% $  
		\end{tabular}
	}
\end{table}

\begin{table}[!htp]
	\centering
	\caption{Nystr\"om (classical) regularised Stokeslet convergence results for comparison purposes: Straight line distance traveled by a single biflagellate swimmer, as described in \S\ref{sec:biflagellate}, after one complete flagellar beat cycle with the Nystr\"om discretisation. The number of points have been chosen following the convergence algorithm in table \ref{tab:heuristic}, with the regularisation parameter $ \epsilon = 0.25/20 $ being the ratio between flagellar radius and length. (a) Distance traveled by the swimmer, (b) and (c) show the percentage change in this distance when doubling $ N^H $ and $N^F$ respectively.}
	\label{tab:chlamyNys}
	\subtable[\ Distance traveled in multiples of $ \left(\text{flagellar length}\right)\cdot 10^{-2} $]{
		\begin{tabular}{ l| l l l l l}
			 \diagbox{$N^H$}{$N^F$} & $ 40 $    & $ 100 $ & $ 200 $ & $ 400 $\\
			\hline
			$ 96 $                  & $ 6.542 $ & $7.148$ & $7.151$ & $7.165$\\
			$ 600 $                 & $ 5.320 $ & $5.794$ & $5.803$ & $5.819$\\
			$ 2646 $                & $ 5.105 $ 	& $5.554$ & $5.563$ & $5.578$
		\end{tabular}
	}
	
	\subtable[\ Percentage change in distance traveled when doubling $ N^F $]{
		\begin{tabular}{ l| l l l l l}
			 \diagbox{$N^H$}{$N^F$} & $ 100 $ & $ 200 $ & $ 400 $\\
			\hline
			$ 96 $                  & $9.29\%$ & $0.05\%$ & $0.19\%$\\
			$ 600 $                 & $8.91\%$ & $0.14\%$ & $0.28\%$\\
			$ 2646 $                & $8.79\%$ & $0.16\%$ & $0.27\%$
		\end{tabular}
	}\qquad\qquad	
	\subtable[\ Percentage change in distance traveled when doubling $ N^H $]{
		\begin{tabular}{ l| l l l l l}
			 \diagbox{$N^H$}{$N^F$} & $ 40 $    & $ 100 $ & $ 200 $ & $ 400 $\\
			\hline
			$ 600 $                 & $18.68\%$ & $18.94\%$ & $18.86\%$ & $18.79\%$\\
			$ 2646 $                & $4.04\%$  & $4.15\%$  & $4.13\%$ & $4.15\%$
		\end{tabular}
	}\end{table}
	
\begin{figure}
  \centering
   \includegraphics[width=\textwidth]{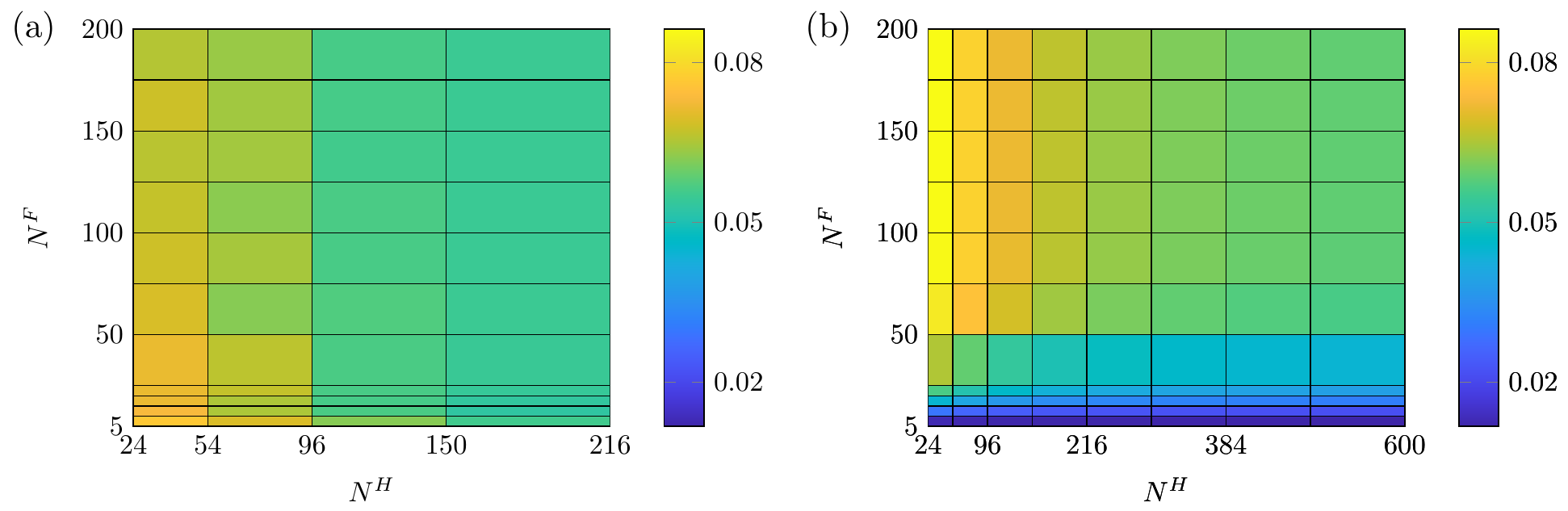}
   \caption{\footnotesize{Visual comparison of how the swimming distance for the biflagellate swimmer converges with (a) the nearest-neighbour and (b) the classic (Nystr\"{om}) discretisations. The swimming distance is shown with increasing number of points for both the head ($N^H$) and flagellum ($N^F$) discretisations. In (a) the quadrature discretisation is chosen to be twice as fine as the force discretisation.}}
   \label{fig:conv}
\end{figure}

\begin{table}[!htb]
	\centering\scriptsize
	\caption{New nearest-neighbour convergence results: Straight line distance traveled by a single sperm, swimming between two opposed surfaces, as described in \S\ref{sec:multiple}, after one complete flagellar beat cycle. Here, the discretisation for the cell head is fixed with $ N^H = 96 $ and $ Q^H = 600 $ force and quadrature points respectively. The number of points discretising the flagellum have been chosen following the convergence algorithm in table \ref{tab:heuristic}, with the regularisation parameter $ \epsilon = 0.25/45 $ being the ratio between flagellar radius and length. When $ N^F > Q^F $ a singular linear system is formed, this is denoted by the entry `NaN' (`not-a-number'). (a) Distance traveled by the swimmer, (b) and (c) show the percentage change in this distance when halving $ h_f^F $ and $h_q^F $ respectively.}
	\label{tab:spermBound}
	\subtable[\ Distance traveled in multiples of $ \left(\text{flagellar length}\right)\cdot 10^{-2} $]{
		\begin{tabular}{ l l l l l| l l l l l}
			         &        &          &       & \multicolumn{1}{r|}{$Q^H$}  & $600$     & $600$     & $600$     & $600$     & $600$      \\		
			         &        &          &       & \multicolumn{1}{r|}{$h_q^H$}& $0.006702$  & $0.006702$  & $0.006702$  & $0.006702$  & $0.006702$   \\
			         &        &          &       &  \multicolumn{1}{r|}{$Q^F$} & $100$     & $200$     & $400$     & $800$     & $1600$     \\
			 DOF     & $N^H$  & $h_f^H$  & $N^F$ & \diagbox{$h_f^F$}{$ h_q^F$} & $0.01011$ & $0.005032$ & $0.002510$ & $0.001254$ & $0.0006265$ \\
			\hline
			$ 408 $ & $ 96 $ & $0.01314$ & $ 40 $  & $ 0.02564 $               & $ 6.912 $ & $ 6.957 $ & $ 6.992 $ & $ 7.004 $ & $ 7.011 $  \\
			$ 528 $ & $ 96 $ & $0.01314$ & $ 80 $  & $ 0.01267 $               & $ 7.003 $ & $ 6.935 $ & $ 6.939 $ & $ 6.749 $ & $ 6.950 $  \\
			$ 768 $ & $ 96 $ & $0.01314$ & $ 160 $ & $ 0.006297 $              & NaN       & $ 6.925 $ & $ 6.924 $ & $ 6.924 $ & $ 6.924 $  \\
			$ 1248 $ & $ 96 $ & $0.01314$ & $ 320 $ & $ 0.003140 $             & NaN       & NaN       & $ 6.920 $ & $ 6.918 $ & $ 6.919 $
		\end{tabular}
	}
	
	\subtable[\ Percentage change in distance traveled when halving $ h_f^F $]{
		\begin{tabular}{ l | l l l l l l}
			\diagbox{$h_f^F$}{$ h_q^F$} & $0.01011$ & $0.005032$ & $0.002510$ & $0.001254$ & $0.0006265$ \\
			\hline
			$ 0.01267 $                  & $ 1.32\%$ & $ 0.31\% $  & $ 0.75\% $  & $ 3.64\% $ & $ 0.86\% $ \\
			$ 0.006297 $                 &           & $ 0.15\% $  & $ 0.21\% $  & $ 2.59\% $ & $ 0.39\% $ \\
			$ 0.003140 $                 &           &             & $ 0.05\% $  & $ 0.09\% $ & $ 0.07\% $
		\end{tabular}
	}	\qquad\qquad	
	\subtable[\ Percentage change in distance traveled when halving $ h_q^F $]{
		\begin{tabular}{ l | l l l l l}
			\diagbox{$h_f^F$}{$ h_q^F$} & $0.005032$ & $0.002510$ & $0.001254$ & $0.0006265$ \\
			\hline
			$ 0.02564 $                  & $ 0.65\% $ & $ 0.50\% $  & $ 0.18\% $ & $ 0.10\% $ \\
			$ 0.01267 $                  & $ 0.97\% $ & $ 0.06\% $  & $ 2.74\% $ & $ 2.98\% $ \\
			$ 0.006297 $                 &            & $ 0.01\% $  & $ 0.01\% $ & $ 0.00\% $ \\
			$ 0.003140 $                 &            &             & $ 0.04\% $ & $ 0.01\% $ 
		\end{tabular}
	}
\end{table}

We have analysed the convergence of the results for the following cases: a single swimming biflagellate (as described in \S\ref{sec:biflagellate}), a single swimming sperm (as in \S\ref{sec:sperm}) with no boundary, and a single swimming sperm with boundary. We also assess the effect of the boundary through fixing the sperm discretisation and applying the heuristic of table \ref{tab:heuristic} to the boundary discretisations, and through fixing the sperm and boundary discretisations and increasing the boundary length. The effects of refining the flagellum discretisations in the biflagellate and single sperm models are shown in tables \ref{tab:chlamy1} and \ref{tab:spermBound}, with the full convergence results provided in the supplemental material. Here, we have used the straight line distance travelled by the swimmer as the objective for convergence, and it is clear from tables \ref{tab:chlamy1} and \ref{tab:spermBound}, together with the associated tables in the supplemental material, that the method is well converged for each swimmer, both in the presence of boundaries and not. Increasing the size of the boundaries resulted in a negligible change to the distance traveled by the swimmer.  We note here that the head discretisation for the sperm case is very fine, this has been chosen to illustrate the convergence results following the heuristic of Smith \cite{smith2018}.

For comparison with our method, in table \ref{tab:chlamyNys} we present the straight line distance travelled by the biflagellate swimmer when the Nystr\"om discretisation is used. We can see from the data in the table \ref{tab:chlamyNys} that the Nystr\"om discretisation requires $8538$ degrees of freedom ($ N^H = 2646 $ and $ N^F = 100 $) to approach within $ 1\% $ the converged distance of $ \approx 5.5\cdot 10^{-2} $ flagellar lengths, while the current method is within $ 1\% $ of the converged distance in the first entry of table \ref{tab:chlamy1}, with only $528$ degrees of freedom ($ N^H = 96 $ and $ N^F = 40 $). In figure \ref{fig:conv} we show the convergence of the swimming distance for both the nearest-neighbour and classic (Nystr\"{o}m) discretisations, where for the former we have chosen the quadrature discretisation to be twice as fine as the force. This figure visually emphases the convergnece results of tables \ref{tab:chlamy1} and \ref{tab:chlamyNys} from which we see that, for the choice of $ \epsilon = 0.25/20$, the Nystr\"om method requires many more degrees of freedom to reach the same levels of convergence. This convergence rate could be improved in the Nystr\"om case by varying $ \epsilon $ (as discussed in \cite{cortez2005}), however as previously discussed the nearest neighbour discretisation is much more robust to this parameter.

\section{Discussion}\label{sec:discussion}
This report has described an extension of the nearest-neighbour regularized Stokeslet method \cite{smith2018} to enable the simulation of multiple force- and moment-free cells swimming in a bounded domain. Cell trajectory calculations were achieved by casting the task as an initial-value problem; by integrating the force at each step it was additionally possible to store the evolving force distribution to enable post-calculation of the velocity field. The method was assessed on two problems of a type which may be of interest in the biological fluid mechanics community: swimming of a biflagellate in an unbounded domain, and motility of multiple human sperm between two no-slip surfaces.

Numerical experiments provide evidence that the method is relatively efficient and converges well, requiring minutes to solve the problems described above, without specialist computational hardware, and we note with interest the significantly improved convergence of this method when compared to the classic Nystr\"om discretisation. While the construction of the matrices is somewhat tedious, the underlying concept of the method --- a coarse/fine discretisation of the boundary integral equations to address the fact that the force distribution varies more slowly than the kernel --- should ensure that the method is comprehensible and extensible by non-specialists. Crucially, no true `mesh' generation (i.e.\ with connectivity tables) is required to simulate a new swimmer of interest. We hope that these properties of ease-of-use, extensibility and efficiency make the method appealing to potential users, and in support of this aim we provide all \matlab code used to generate this report in the repository \texttt{github.com/djsmithbham/nearestStokesletSwimmers}. Within this repository, a template file \texttt{nnSwimmerTemplate.m} is provided which sets out how new swimmers can be added to the existing codebase. 

There are many potential extensions for this work spanning the whole field of locomotion at low Reynolds number. The convergence properties of this method mean that it may be valuable for high-throughput analysis of experimental data, or (perhaps with adaptations to deal efficiently with long-range interactions) suspensions of relatively large numbers of swimmers. It would be interesting to see if the modification of the method to take into account viscoelastic effects would allow for the collective swimming behaviour of sperm seen by Tung et al. \cite{tung2017} to be reproduced from an idealised model of swimming. There is potential for this method to be applied to the world of phoretic swimmers to examine the dynamics of many phoretic particles, or to the case of swimmers driven by magnetic fields. The computational efficiency of this method can also be exploited through modelling multiple swimmers in complex environments, for example ciliary flow. While such flows would previously have been simulated and then applied as a background flow to a swimmer, with this efficient method one would be able to model the ciliary beating patterns directly and could allow for a more realistic interaction between swimmers and their environment.

\section{Acknowledgements}
This work was supported by Engineering and Physical Sciences Research Council award EP/N021096/1. We would also like to thank Gemma Cupples (University of Birmingham) for helpful comments on the manuscript and template files in the code repository, and Marco Polin (University of Warwick), Hermes Gad{\^e}lha (University of York), Eamonn Gaffney and Kenta Ishimoto (University of Oxford), and Hao Wu (University of Minnesota) for helpful discussions.

See Supplemental Material at [URL will be inserted by publisher] for the complete set of convergence tables for the swimmers provided in this manuscript.

\appendix

\bibliographystyle{unsrt}

\begin{thebibliography}{11}

\bibitem{taylor1951}
G.I. Taylor.
\newblock Analysis of the swimming of microscopic organisms.
\newblock {\em Proc. R. Soc. Lond. A.}, 209:447--461, 1951.

\bibitem{keaveny2013}
E.E. Keaveny, S.W. Walker, and M.J. Shelley.
\newblock Optimization of chiral structures for microscale propulsion.
\newblock {\em Nano Lett.}, 13(2):531--537, 2013.

\bibitem{simons2015}
J.~Simons, L.~Fauci, and R.~Cortez.
\newblock {A fully three-dimensional model of the interaction of driven elastic
  filaments in a Stokes flow with applications to sperm motility}.
\newblock {\em J. Biomech.}, 48(9):1639--1651, 2015.

\bibitem{ishimoto2017}
K.~Ishimoto, H.~Gad{\^e}lha, E.A. Gaffney, D.J. Smith, and J.~Kirkman-Brown.
\newblock Coarse-graining the fluid flow around a human sperm.
\newblock {\em Phys. Rev. Lett.}, 118(12):124501, 2017.

\bibitem{gray1955}
J.~Gray and G.J. Hancock.
\newblock The propulsion of sea urchin spermatozoa.
\newblock {\em J. Exp. Biol.}, 32:802--814, 1955.

\bibitem{omalley2012}
S.~O'{M}alley and M.~A. Bees.
\newblock The orientation of swimming biflagellates in shear flows.
\newblock {\em B. Math. Biol.}, 74(1):232--255, 2012.

\bibitem{montenegro201}
T.~D. Montenegro-Johnson, L.~Koens, and E.~Lauga.
\newblock Microscale flow dynamics of ribbons and sheets.
\newblock {\em Soft matter}, 13(3):546--553, 2017.

\bibitem{huang2017}
J.~Huang and L.~Fauci.
\newblock Interaction of toroidal swimmers in stokes flow.
\newblock {\em Phys. Rev. E}, 95(4):043102, 2017.

\bibitem{schmieding2017}
L.~C. Schmieding, E.~Lauga, and T.~D. Montenegro-Johnson.
\newblock Autophoretic flow on a torus.
\newblock {\em Phys. Rev. Fluids}, 2(3):034201, 2017.

\bibitem{montenegro2018}
T.~D. Montenegro-Johnson.
\newblock Microtransformers: controlled microscale navigation with flexible
  robots.
\newblock {\em arXiv preprint arXiv:1801.09742v1}, 2018.

\bibitem{tung2017}
C.-k. Tung, C.~Lin, B.~Harvey, A.~G. Fiore, F.~Ardon, M.~Wu, and S.~S. Suarez.
\newblock Fluid viscoelasticity promotes collective swimming of sperm.
\newblock {\em Sci. Rep.}, 7(1):3152, 2017.

\bibitem{cripe2016}
P.~Cripe, O.~Richfield, and J.~Simons.
\newblock Sperm pairing and measures of efficiency in planar swimming models.
\newblock {\em Spora A. J. Biomath.}, 2(1):5, 2016.

\bibitem{simons2014}
J.~Simons, S.~Olson, R.~Cortez, and L.~Fauci.
\newblock The dynamics of sperm detachment from epithelium in a coupled
  fluid-biochemical model of hyperactivated motility.
\newblock {\em J. Theor. Biol.}, 354:81--94, 2014.

\bibitem{lushi2017}
E.~Lushi, V.~Kantsler, and R.~E. Goldstein.
\newblock Scattering of biflagellate microswimmers from surfaces.
\newblock {\em Phys. Rev. E}, 96(2):023102, 2017.

\bibitem{shum2017}
H.~Shum and J.~M. Yeomans.
\newblock Entrainment and scattering in microswimmer-colloid interactions.
\newblock {\em Phys. Rev. Fluids}, 2(11):113101, 2017.

\bibitem{zottl2017}
A.~Z{\"o}ttl and J.~M. Yeomans.
\newblock Enhanced bacterial swimming speeds in macromolecular polymer
  solutions.
\newblock {\em arXiv preprint arXiv:1710.03505}, 2017.

\bibitem{pedley1990}
T.~J. Pedley and J.~O. Kessler.
\newblock A new continuum model for suspensions of gyrotactic micro-organisms.
\newblock {\em J. Fluid Mech.}, 212:155--182, 1990.

\bibitem{gaffney2011}
E.~A. Gaffney, H.~Gad{\^e}lha, D.~J. Smith, J.~R. Blake, and J.~C.
  Kirkman-Brown.
\newblock Mammalian sperm motility: observation and theory.
\newblock {\em Annu. Rev. Fluid Mech.}, 43:501--528, 2011.

\bibitem{brokaw1971}
C.~J. Brokaw.
\newblock Bend propagation by a sliding filament model for flagella.
\newblock {\em J. Exp. Biol.}, 55(2):289--304, 1971.

\bibitem{higdon1979}
J.J.L. Higdon.
\newblock A hydrodynamic analysis of flagellar propulsion.
\newblock {\em J. Fluid Mech.}, 90:685--711, 1979.

\bibitem{phan1987}
N.~Phan-Thien, T.~Tran-Cong, and M.~Ramia.
\newblock A boundary-element analysis of flagellar propulsion.
\newblock {\em J. Fluid Mech.}, 185:533--549, 1987.

\bibitem{cortez2005}
R.~Cortez, L.~Fauci, and A.~Medovikov.
\newblock The method of regularized {S}tokeslets in three dimensions:
  {A}nalysis, validation, and application to helical swimming.
\newblock {\em Phys. Fluids}, 17:031504, 2005.

\bibitem{gillies2009}
E.A. Gillies, R.M. Cannon, R.B. Green, and A.A. Pacey.
\newblock Hydrodynamic propulsion of human sperm.
\newblock {\em J. Fluid Mech.}, 625:445, 2009.

\bibitem{shankar2015}
V.~Shankar and S.~D. Olson.
\newblock Radial basis function (rbf)-based parametric models for closed and
  open curves within the method of regularized Stokeslets.
\newblock {\em Int. J. Numer. Meth. Fl.}, 79(6):269--289, 2015.

\bibitem{rostami2016}
M.~W. Rostami and S.~D Olson.
\newblock Kernel-independent fast multipole method within the framework of
  regularized Stokeslets.

\bibitem{schoeller2018}
S.~F. Schoeller and E.~E. Keaveny.
\newblock From flagellar undulations to collective motion: predicting the
  dynamics of sperm suspensions.
\newblock {\em arXiv preprint arXiv:1801.08180}, 2018.

\bibitem{smith2009}
D.J. Smith, E.A. Gaffney, J.R. Blake, and J.C. Kirkman-Brown.
\newblock Human sperm accumulation near surfaces: a simulation study.
\newblock {\em J. Fluid Mech.}, 621:289--320, 2009.

\bibitem{peskin2002}
C.~S. Peskin.
\newblock The immersed boundary method.
\newblock {\em Acta Numer.}, 11:479--517, 2002.

\bibitem{li2017}
C.~Li, B.~Qin, A.~Gopinath, P.~E. Arratia, B.~Thomases, and R.~D. Guy.
\newblock Flagellar swimming in viscoelastic fluids: role of fluid elastic
  stress revealed by simulations based on experimental data.
\newblock {\em J. R. Soc. Interface}, 14(135):20170289, 2017.

\bibitem{smith2018}
D.J. Smith.
\newblock {A nearest-neighbour discretisation of the regularized Stokeslet
  boundary integral equation}.
\newblock {\em J. Comput. Phys.}, 358:88--102, 2018.

\bibitem{liron1981}
N.~Liron and J.R. Blake.
\newblock Existence of viscous eddies near boundaries.
\newblock {\em J. Fluid Mech.}, 107:109--129, 1981.

\bibitem{ainley2008}
J.~Ainley, S.~Durkin, R.~Embid, P.~Boindala, and R.~Cortez.
\newblock The method of images for regularized {S}tokeslets.
\newblock {\em J. Comput. Phys.}, 227:4600--4616, 2008.

\bibitem{sartori2016}
P.~Sartori, V.F. Geyer, A.~Scholich, F.~J{\"u}licher, and J.~Howard.
\newblock Dynamic curvature regulation accounts for the symmetric and
  asymmetric beats of chlamydomonas flagella.
\newblock {\em e{L}ife}, 5, 2016.

\bibitem{dresdner1981}
R.D. Dresdner and D.F. Katz.
\newblock Relationships of mammalian sperm motility and morphology to
  hydrodynamic aspects of cell function.
\newblock {\em Biol. Reprod.}, 25(5):920--930, 1981.

\end{thebibliography}

\end{document}